\documentclass{elsarticle}
\usepackage{latexsym}
\usepackage{longtable}
\usepackage{float}
\usepackage[latin2]{inputenc}
\usepackage{graphicx}
\usepackage{ae}
\usepackage{aecompl}
\usepackage{setspace}
\usepackage{amsmath}
\usepackage{amssymb}
\usepackage{amsthm}
\usepackage[hidelinks]{hyperref}

\makeatletter
\def\ps@pprintTitle{%
	\let\@oddhead\@empty
	\let\@evenhead\@empty
	\def\@oddfoot{\centerline{\thepage}}%
	\let\@evenfoot\@oddfoot}
\makeatother

\begin{document}

\title{The Frequent Network Neighborhood Mapping of the Human Hippocampus Shows Much More Frequent Neighbor Sets in Males Than in Females }
	
\author[p]{Máté Fellner}
\ead{fellner@pitgroup.org}
\author[p]{Bálint Varga}
\ead{balorkany@pitgroup.org}
\author[p,u]{Vince Grolmusz\corref{cor1}}
\ead{grolmusz@pitgroup.org}
\cortext[cor1]{Corresponding author}
\address[p]{PIT Bioinformatics Group, Eötvös University, H-1117 Budapest, Hungary}
\address[u]{Uratim Ltd., H-1118 Budapest, Hungary}

\date{}

\begin{abstract}
		In the study of the human connectome, the vertices and the edges of the network of the human brain are analyzed: the vertices of the graphs are the anatomically identified gray matter areas of the subjects; this set is exactly the same for all the subjects. The edges of the graphs correspond to the axonal fibers, connecting these areas. In the biological applications of graph theory, it happens very rarely that scientists examine numerous large graphs on the very same, labeled vertex set. Exactly this is the case in the study of the connectomes. Because of the particularity of these sets of graphs, novel, robust methods need to be developed for their analysis. Here we introduce the new method of the Frequent Network Neighborhood Mapping for the connectome, which serves as a robust identification of the neighborhoods of given vertices of special interest in the graph. We apply the novel method for mapping the neighborhoods of the human hippocampus and discover strong statistical asymmetries between the connectomes of the sexes, computed from the Human Connectome Project. We analyze 413 braingraphs, each with 463 nodes.  We show that the hippocampi of men have much more significantly frequent neighbor sets than women; therefore, in a sense, the connections of the hippocampi are more regularly distributed in men and more varied in women. Our results are in contrast to the volumetric studies of the human hippocampus, where it was shown that the relative volume of the hippocampus is the same in men and women.
\end{abstract}

\maketitle
	
\section*{Introduction} While it seems to be clear for all brain scientists that the complex connection patterns of the neurons play a fundamental role in brain function \cite{Seung2009,Sporns2005,Lichtman2008a}, when the large-scale, macroscopic description of these connections has become available by the development of diffusion MRI techniques, it turned out that novel methods are needed to handle these large graphs \cite{Seung2009,Sporns2005}. MRI-mapped human connectomes have only several hundred or at most one thousand vertices today \cite{Hagmann2012}, and, therefore, more complex, more refined graph theoretical algorithms \cite{Szalkai2015,Kerepesi2015b,Szalkai2015c} can be applied for their analysis than the widely followed network science approach, originally developed for tens of millions of vertices of the web graph \cite{Faloutsos1999}.

\subsection*{The need for robust analytical methods}

In the mathematical analysis of the large biological graphs, it is necessary to apply methods that are capable of dealing with the frequently erroneous networks, computed from high-deviation experimental data \cite{Goll2006,Gavin2006,Krogan2006}. Our research group has developed several fault-tolerant analytical methods for biological graphs, based on a well-known robust network measure: Google's PageRank \cite{Grolmusz2015a,Grolmusz2015,Banky2013,Ivan2011}. 

Today, one of the main challenges in brain science is the detailed mapping of the brain circuitry with reliable methods. While diffusion MRI yields a relatively reliable set of tools for brain imaging, the workflow of identifying the common, human brain areas across different subjects and the tractography phase of the processing have numerous difficulties \cite{Zhan2015,Jbabdi2011,Bastiani2012}. Human braingraphs, or connectomes are very focused structures for the description of the cerebral connections: the nodes of these graphs are the anatomically identified small (1-1.5 cm$^2$) areas of the gray matter (called ROIs, Regions of Interests), and two nodes, corresponding to two ROIs, are connected by an edge if in the tractography phase an axonal fiber is found, connecting these two ROIs. These connectomes describe the macroscopical scale set of connections between hundreds of brain areas. Before the era of diffusion MR imaging, only very fractional knowledge was available on these connections \cite{Seung2009}; today tens of thousands of connections can be identified and examined.

\subsection*{Previous Work}

Perhaps the most straightforward robust approach to be considered is the study of the frequently appearing cerebral connections. In work \cite{Kerepesi2015a} we have mapped the differences in the individual variability of the connections within the lobes and some smaller brain areas.

We have constructed the Budapest Reference Connectome Server \cite{Szalkai2015a,Szalkai2016} at the address \url{https://pitgroup.org/connectome/}, which is capable of generating consensus connectomes from the data of 477 subjects, consisting of $k$-frequent edges (i.e., edges that are present in at least $k$ braingraphs), with user-selected $k$ and other parameters. 

The Budapest Reference Connectome Server is an excellent tool for generating a robust human connectome, and, additionally, its construction has led to the discovery of the phenomenon of the Consensus Connectome Dynamics (CCD) \cite{Kerepesi2016,Szalkai2016e,Kerepesi2015b,Szalkai2016d}, mirroring the development of the axonal connections within the human brain.

The global, or general approaches for describing the frequent (and, therefore, robust) cerebral connections are not always detailed enough for the study of specific brain regions. Additionally, the frequently appearing connections (e.g., in the Budapest Reference Connectome Server \cite{Szalkai2015a,Szalkai2016}) may describe the frequent neighbors of the individual vertices of these graphs, but not the frequently appearing neighbor-sets of important vertices. The description of these neighbor-sets is the goal of the present work.

\subsection*{Our contribution: the Frequent Network Neighborhood Mapping}

Let us consider an important small area, corresponding to a vertex in the connectome (called ROI, region of interest) of the brain, say the left hippocampus. It is an important question to describe those ROIs, which are directly connected to the left hippocampus since all the connections to and from the other cerebral areas go through these edges and these neighbors of the important ROI (in our case the hippocampus). It is a very interesting question whether almost all the subjects have the same neighbors of the hippocampus, or there is a considerable variability among the subjects.

If there were no any variability among the connectomes of the individual subjects, then in each connectome, the left hippocampus would be connected to the very same set of other nodes or ROIs. However, there is a considerable variability of these connections between distinct subjects \cite{Kerepesi2015a,Szalkai2015a,Szalkai2016}. Therefore, no such common neighbor-set exists for any vertex in the braingraphs. 

Instead of the non-existent single, universal neighbor-set, which would have been present in the connectome of all the subjects, we can still identify at least the frequently appearing neighbor-sets of the left hippocampus (or any other given vertex of interest in the graph). 

Clearly, this is a completely different task than identifying the frequently appearing edges of the connectome, which are mapped by the Budapest Reference Connectome Server at \url{https://pitgroup.org/connectome}. As an example, let us consider a vertex $u$, and two other vertices $v$ and $w$. Suppose, that both edges $\{u,v\}$ and $\{u,w\}$ are present in 90\% of all connectomes, but it may happen that the vertex-set $\{v,w\}$ appears only in the 80\% of the graphs as the neighbors of vertex $v$: if the connectomes are indexed from 1 through 100, it may happen that in connectomes 1 through 90 $\{u,v\}$ is an edge; in connectomes 11 through 100 $\{u,w\}$ is an edge; so both of them are present in 90\% of the connectomes, but the set $\{v,w\}$ is neighboring with $u$ from connectomes 11 through 90, i.e., only 80 connectomes, i.e., 80\%. 

In the present contribution, we map the frequently appearing neighbor-sets of the left and the right hippocampi of the human connectome, and we make comparisons between the lateral and sex-differences in the frequent neighbor-sets of the left- and right hippocampi. The neighbor set sizes are bounded by 4 in our present study, since if we considered larger sets than 4, the numbers of the sets would be increased considerably. Our braingraphs have 463 vertices for all subjects.

Our results show strong differences in the neighbor-sets of the hippocampi between the sexes: we mapped the neighbor-sets, which have significant differences in frequency in men and women (we call these vertex sets ``significant neighbor-sets''), and we have found that
\begin{itemize}
\item the number of the significant neighbor-sets of the left hippocampus is 65 times higher in males than in females; 

\item the number of the significant neighbor-sets of the right hippocampus is 16 times higher in males than in females;
\end{itemize}

In a sense, these results show that the neighbor sets of the hippocampus of the women are more varied between individuals, while these sets are more regular, that is, less varied in the case of men's connectomes. These results complement our general studies of the deep graph-theoretical parameters of the connectomes of men and women \cite{Szalkai2015,Szalkai2016a,Szalkai2015c}, where it was proven that women's connectomes are better ``connected'', in precisely defined mathematical and computer engineering terms.

\section*{Discussion and Results}

Mapping the frequent graph-theoretical structures in the human braingraph makes possible the robust analysis of the possibly noisy data. If the frequency count is set to a high enough value, (say, the structure in question needs to be present in 80\% or 90\% of the connectomes of the subjects in the group analyzed), then image acquisition artifacts, data processing errors and small, random individual variabilities in the connectomes could be counter-measured.

Here we consider the hippocampus and its neighbors of the human brain. The hippocampus plays an important role in numerous brain functions, like the processing of short-time memory and turning it into long-time memory, in spatial memory and orientation \cite{Santin2000,Voineskos2015,Nees2014,Iglesias2015}. Today, the hippocampus is perhaps the most widely studied functional and structural entity of the human brain, and, consequently, the detailed study of its neighbor sets is a relevant area. Additionally, the detailed study of the cerebral circuitry is an emphasized research topic today: describing the direct neighbors of one of the most important brain areas in this sense is also a crucial question.

In our present study, we have discovered and analyzed frequent 1,2,3 and 4 element neighbor-sets of the left and the right human hippocampi.

\subsection*{Frequent Network Neighborhood Mapping of the hippocampus}

From now on, we refer to the brain areas, or ROIs, by their resolution-250 parcellation labels, based on the Lausanne 2008 brain atlas \cite{Hagmann2008} and computed by using FreeSurfer \cite{Fischl2012} and CMTK \cite{Gerhard2011,Daducci2012}, listed at 
\url{https://github.com/LTS5/cmp_nipype/blob/master/cmtklib/data/parcellation/lausanne2008/ParcellationLausanne2008.xls}. The ``lh'' and the ``rh'' prefixes abbreviate the ``left-hemisphere'' and ``right-hemisphere'' localizations.

We have mapped separately the direct neighbor sets of the left- and the right hippocampi. 

We need to recall an important property of the frequencies of the neighbor sets of a given vertex, in our case the left- or the right hippocampus. We say that a set $U$ is a subset of set $V$, if every element of $U$ is, at the same time, the element of $V$. $V$ is called the superset of $U$. We denote this relation as follows: $U\subseteq V$ or, equivalently, $V\supseteq U$. 

Let us consider the left hippocampus. For every neighbor-set $U$ of the left hippocampus, we assign a frequency value as follows: we count the graphs of the subjects, where every element of set $U$ is connected to the left hippocampus, and divide this number by the number of all the graphs. Let us denote this frequency value by $\phi(U)$. Clearly, if $V$ is a superset of $U$, then the frequency of $V$ cannot be larger than the frequency of $U$:  $\phi(U)\geq \phi(V)$.

This observation concerning the frequencies, of course, holds for the neighbor-sets of any vertex of our graphs.

\begin{table}
\centering
\scriptsize
\begin{tabular}{ l l }
	Frequency & Subset \\
1	&	(Left-Putamen)\\
1	&	(Left-Thalamus-Proper)\\
1	&	(lh.isthmuscingulate\_3)\\
1	&	(Left-Putamen)(Left-Thalamus-Proper)\\
1	&	(Left-Putamen)(lh.isthmuscingulate\_3)\\
1	&	(Left-Thalamus-Proper)(lh.isthmuscingulate\_3)\\
1	&	(Left-Putamen)(Left-Thalamus-Proper)(lh.isthmuscingulate\_3)\\
.   &    ...........................................................\\
0.99758	&	(lh.superiortemporal\_2)\\
0.99758	&	(Left-Putamen)(lh.superiortemporal\_2)\\
0.99758	&	(Left-Thalamus-Proper)(lh.superiortemporal\_2)\\
0.99758	&	(lh.isthmuscingulate\_3)(lh.superiortemporal\_2)\\
0.99758	&	(Left-Putamen)(Left-Thalamus-Proper)(lh.superiortemporal\_2)\\
0.99758	&	(Left-Putamen)(lh.isthmuscingulate\_3)(lh.superiortemporal\_2)\\
0.99758	&	(Left-Thalamus-Proper)(lh.isthmuscingulate\_3)(lh.superiortemporal\_2)\\
0.99758	&	(Left-Putamen)(Left-Thalamus-Proper)(lh.isthmuscingulate\_3)(lh.superiortemporal\_2)\\
....... &   ...................................................................................\\
0.99517	&	(lh.insula\_2)\\
0.99517	&	(Left-Putamen)(lh.insula\_2)\\
0.99517	&	(Left-Thalamus-Proper)(lh.insula\_2)\\
0.99517	&	(lh.insula\_2)(lh.isthmuscingulate\_3)\\
0.99517	&	(Left-Putamen)(Left-Thalamus-Proper)(lh.insula\_2)\\
0.99517	&	(Left-Putamen)(lh.insula\_2)(lh.isthmuscingulate\_3)\\
0.99517	&	(Left-Thalamus-Proper)(lh.insula\_2)(lh.isthmuscingulate\_3)\\
0.99517	&	(Left-Putamen)(Left-Thalamus-Proper)(lh.insula\_2)(lh.isthmuscingulate\_3)\\

\end{tabular}
\caption{The most frequent subsets of the neighbors of the left hippocampus (cut frequency value: 0.995). 413 subjects from the Human Connectome Project public release were examined. In all the subjects, the left hippocampus is connected to the following three ROIs: the Left-Putamen, Left-Thalamus-Proper and the lh.isthmuscingulate\_3, and, therefore, to all 7 ($=2^3-1$) non-empty subsets of those (the first 7 lines of the table). The lh.superiortemporal\_2 ROI is connected to the left hippocampus in all, but one subjects, so its frequency is 412/413=0.99758. Note that no subset, containing lh.superiortemporal\_2, may have a higher frequency than the frequency of lh.superiortemporal\_2 alone, i.e., 0.99758. Indeed, all the 8 (empty and non-empty) subsets of the Left-Putamen, Left-Thalamus-Proper and the lh.isthmuscingulate\_3, together with the lh.superiortemporal\_2 is present in 412 out of 413 subjects, i.e., with 0.99517 frequency. The ROI lh.insula\_2 is connected to the left hippocampus in 411 subjects out of the 413 subjects, therefore its frequency is 411/413=0.99517. Similarly, all the 8 subsets of the ROIs Left-Putamen, Left-Thalamus-Proper, and the lh.isthmuscingulate\_3, together with lh.insula\_2 have the same 411/413=0.99517 frequency. Theoretically, lh.insula\_2, together with the lh.superiortemporal\_2 may have the same frequency as lh.insula\_2 alone (if the subject, where the left hippocampus - lh.superiortemporal\_2 edge is missing is one of the two subjects where the left hippocampus - lh.insula\_2 edge is missing), but this is not the case: the frequency of the neighbor set \{lh.superiortemporal\_2,lh.insula\_2\} is below the cut frequency value for Table 1, i.e., 0.995.}
\end{table}

Table 1 lists the neighbor-sets of the left hippocampus with a minimum frequency of 0.995. Three ROIs (the Left-Putamen, Left-Thalamus-Proper and the lh.isthmuscingulate\_3) are connected to the left hippocampus in all braingraphs, while several other ROI (such as the lh.superiortemporal\_2 and the lh.insula\_2), together with some subsets of the first three ROIs form the remaining neighbor sets with a minimum frequency of 0.995.

The supporting Tables S1 and S2 contain the one, two three and four-element subsets of the neighbor-sets of the left- and the right hippocampi, respectively, with a minimum frequency of 0.9. 

\subsection*{Sex differences}

In what follows we compare the neighbor sets of the left- and right hippocampi in braingraphs, computed from the male and female subjects of the dataset of the Human Connectome Project \cite{McNab2013}. We identify those neighbor sets of the hippocampus that are significantly more frequent in male- and in female connectomes. We will see that male connectomes have much more frequent hippocampal neighbor sets than female connectomes. 

By our knowledge, this is the first observed significant sex difference in the connections of the hippocampus in the literature.

Anatomical sex differences in the volume of the hippocampus were studied in \cite{Tan2016}: it was found that males have larger absolute hippocampus volumes, but the relative hippocampus volume, compared to both the total brain volume or the intracranial volume, are the same in the two sexes.

Here we show that there are 65 times more frequent neighbor sets of size at most 4 of the left hippocampus in males than in females; and there are 16 times more 
frequent neighbor sets of size at most 4 of the right hippocampus in males than in females. These results show that the variability of the neighbor sets of the hippocampus is smaller in the case of males than in females: in males there are much more frequent neighbor sets than in females. In other words, the variability of the neighbor sets of the hippocampus in the connectomes of women is greater than in the case of men.

\begin{table}
	\scriptsize
\begin{tabular}{ | c | l | c | r | r | r | r | c | c | }
	\hline

	&  &  & Size & Size & Size & Size   &\# significant \  &Sign.\ diff.\ \  \\ \hline 
	Support & Node & sex & 1 & 2 &  3 & 4 &  differences\ &  for whom \\ \hline \hline
	0.8 & HPC left & male & 45 & 844 & 9102 & 65150 & 15732 & 15497 \\ \hline
	0.8 & HPC left & female & 38 & 679 & 7128 & 48717 &  & 235 \\ \hline
	0.8 & HPC right & male & 50 & 1023 & 11989 & 91498 & 1762 & 1659 \\ \hline
	0.8 & HPC right & female & 50 & 997 & 10725 & 73424 &  & 103 \\ \hline
	0.9 & HPC & male & 64 & 1781 & 28823 & 309659 & 19828 & 17688 \\ \hline
	0.9 & HPC & female & 62 & 1406 & 17976 & 150167 &  & 2140 \\ \hline
	0.8 & HPC left & all & 43 & 759 & 7863 & 54861 &  &  \\ \hline
	0.8 & HPC right & all & 52 & 1059 & 11963 & 85624 &  &  \\ \hline
	0.9 & HPC & all & 68 & 1728 & 24741 & 233548 &  &  \\ \hline
\end{tabular}
\caption{The summary of the results for sex differences. The first column list the minimum support, or, in other words, the frequency cut-off values: there are two values: 0.8 or 0.9, i.e., 80\% 90\%. The second column denotes the righ-, left- or both hippocampi; the abbreviation HPC stands for the word ``hippocampus''. In the third column the sex is given; the next four columns contain the number size 1, 2, 3 and 4 frequent neighbor-sets of the hippocampus considered. The next column gives the number of the neighbor-sets, which have significantly different frequencies (p=0.001) in male and female connectomes. The last, ninth column gives the number of neighbor-sets, which are significantly more frequent in male or in female connectomes: the sum of the two numbers in the ninth column is equal to the number in the eighth column. 	
	For example, in the first row, we can see that in males, the left hippocampus has 45 frequent 1-element neighbor sets; 844 frequent 2-element neighbor sets, 9102 3-element neighbor sets and 65150 frequent 4-element neighbor sets, where the frequency cut-off is 0.8. Moreover, one can see that there are 15732 sets, differing significantly in frequency in males and in females; and the last column says that from these 15732 sets, 15497 are present in the braingraph of males and only 235 in the braingraphs of females. }
\end{table}

Table 2 summarizes the results for the sex differences in the frequent neighbor sets of the hippocampus. In size 1, 2, 3 and 4 neighbor-sets, males have more frequent sets than females. The only exception is the 1-element frequent neighbors of the right hippocampus, where both males and females have 50 frequent singleton sets. Since all the elements of the frequent 2,3 and 4 element subsets need to be present also as a frequent 1-element set, it is surprising that there is such a big difference in the 4 element frequent neighbors of the right hippocampus in males (91498 sets) and females (73424 sets). 

The following observation is much more surprising: There is a definitive, but not two large difference between the numbers of the frequent size-1,2,3 and 4 neighbor sets of the left and the right hippocampi between the sexes. If we consider, however, the number of neighbor sets with frequencies {\em statistically significantly differing} (with p=0.001) between the two sexes, we have got that 15732 sets differ, and from these, 15497 is significantly more frequent in male-, and 235 is significantly more frequent in female connectomes, in the case of the neighbors of the left hippocampus. Neighbor sets of the right hippocampus have 1762 significant differences, from which 1659 is significantly more frequent in males, and 103 in female connectomes.

If we take the union of the neighbor sets of the left- and right hippocampi, then the number of the neighbor-sets with significant differences is 19828, from which 17688 are more frequent in male connectomes, and 2140 in female connectomes (p=0.001).

Table 3 lists 10 neighbor-sets of the left hippocampus with the most significantly different frequencies in the sexes, where the higher frequency appears at the male )5 sets) and also at the female subjects (5 sets). The supporting Table S3 lists those neighbor-sets of the left hippocampus, which are significantly more frequent in female connectomes; while Table S4 lists those, which are significantly more frequent in male connectomes.

\begin{table}
	\scriptsize
	\centering
\begin{tabular}{ | l | l | l | l | l | }
	\hline
	frequency  &  frequency &  name \\ \hline
	\ male & \ female & \    \\ \hline
	0.823 & 0.556 &   (lh.fusiform\_4)(lh.precuneus\_10)(lh.precuneus\_11)(lh.superiortemporal\_6)\\ \hline
	0.823 & 0.561 &   (lh.fusiform\_4)(lh.isthmuscingulate\_2)(lh.precuneus\_6)(lh.superiortemporal\_6)\\ \hline
	0.823 & 0.561 &   (lh.fusiform\_4)(lh.lingual\_6)(lh.precuneus\_10)(lh.superiortemporal\_6)\\ \hline
	0.823 & 0.561 &   (lh.fusiform\_4)(lh.lingual\_8)(lh.precuneus\_10)(lh.superiortemporal\_6)\\ \hline
	0.823 & 0.561 &   (lh.fusiform\_4)(lh.isthmuscingulate\_2)(lh.precuneus\_10)(lh.superiortemporal\_6)\\ \hline
	\  & \  & \   \\ \hline
	 0.880    &  0.967    &  (Brain-Stem)(lh.bankssts\_2)\\ \hline
	 0.880    &  0.967    &  (Brain-Stem)(Left-Thalamus-Proper)(lh.bankssts\_2)(lh.isthmuscingulate\_3)\\ \hline
	 0.880    &  0.967    &  (Brain-Stem)(Left-Putamen)(lh.bankssts\_2)(lh.isthmuscingulate\_3)\\ \hline
	 0.880    &  0.967    &  (Brain-Stem)(Left-Thalamus-Proper)(lh.bankssts\_2)\\ \hline
	 0.880    &  0.967    &  (Brain-Stem)(lh.bankssts\_2)(lh.isthmuscingulate\_3)\\ \hline
\end{tabular}
\caption{Several neighbor-sets of the left hippocampus with the most significant differences in the frequencies between the sexes. The first five rows list five subsets, which are more frequent in the braingraphs of men than of women. It is interesting that the lh.fusiform\_4 ROI is present in the five sets. The next five are the most significant sets with frequencies higher in females than in males. The lh.bankssts\_2 and the Brain-Stem ROIs are present in all five sets. The supporting Table S3 lists those neighbor-sets of the left hippocampus, which are significantly more frequent in female connectomes; while supporting Table S4 lists those, which are significantly more frequent in male connectomes, $p\leq0.01$. Similarly, Tables S5 and S6 contain the analogous data for the right hippocampus. }
\end{table}

Table 4 lists 10 neighbor-sets of the right hippocampus with the most significantly different frequencies in the sexes, where the higher frequency appears at the male (5 sets) and also at the female subjects (5 sets). The supporting Table S5 lists those neighbor-sets of the right hippocampus, which are significantly more frequent in female connectomes; while Table S6 lists those, which are significantly more frequent in male connectomes. The significance threshold is p=0.01 (corrected by the Holm-Bonferroni method).

\begin{table}
	\scriptsize
	\centering
\begin{tabular}{ | l | l | l | }
	\hline
	frequency  &  frequency  &  name \\ \hline
	\ male & \ female  & \  \\ \hline
	0.840    &  0.682    &  (rh.entorhinal\_1)(rh.parahippocampal\_2)(rh.precuneus\_4)(rh.transversetemporal\_1)\\ \hline
	0.909    &  0.774    &  (rh.entorhinal\_1)(rh.parahippocampal\_2)(rh.precuneus\_2)(rh.precuneus\_4)\\ \hline
	0.914    &  0.782    &  (Brain-Stem)(rh.entorhinal\_1)(rh.parahippocampal\_2)(rh.precuneus\_4) \\ \hline
	0.914    &  0.787    &  (Brain-Stem)(rh.entorhinal\_1)(rh.precuneus\_2)(rh.precuneus\_4)\\ \hline
	0.914    &  0.787    &  (rh.bankssts\_2)(rh.entorhinal\_1)(rh.parahippocampal\_2)(rh.precuneus\_4)\\ \hline
	\  & \  & \  \\ \hline
	0.736 & 0.862 &   (rh.fusiform\_8)(rh.inferiortemporal\_1)(rh.middletemporal\_9) \\ \hline
	0.769 & 0.887 &   (rh.fusiform\_8)(rh.inferiorparietal\_4)(rh.lingual\_7)(rh.middletemporal\_9)\\ \hline
	0.769 & 0.887 &   (rh.fusiform\_8)(rh.middletemporal\_9)(rh.supramarginal\_9) \\ \hline
	0.791 & 0.900 &   (rh.fusiform\_8)(rh.lingual\_7)(rh.middletemporal\_9) \\ \hline
	0.791 & 0.900 &   (Right-Amygdala)(rh.fusiform\_8)(rh.middletemporal\_9) \\ \hline
\end{tabular}
\caption{Several neighbor-sets of the right hippocampus with the most significant differences in the frequencies between the sexes. The first five rows list five subsets, which are more frequent in the braingraphs of men than of women. It is interesting that the rh.precuneus\_4 ROI is present in all five sets. The next five sets are the most significant with frequencies higher in females than in males. The rh.fusiform\_8 and the rh.middletemporal\_9 ROIs are present in all five sets. The supporting Table S5 lists those neighbor-sets of the right hippocampus, which are significantly more frequent in female connectomes; while Table S6 lists those, which are significantly more frequent in male connectomes, $p\leq0.01$.}
\end{table}

As we can see in Table 2, there is not a large difference between the frequent 1-element neighbors of the hippocampus in men and women. However, there is a very significant difference in neighbor-set frequencies of higher cardinality, as it is described in the last two columns of Table 2. One possible reason for this could be that the neighbor sets of the hippocampus in general and the left hippocampus, in particular, have less variability in the case of men than in the case of women: men have the more regularly appearing neighbor-sets, while women have more varied neighbor-sets.

\section*{Materials and Methods}

The primary data source of the present study is Human Connectome Project's website at \url{http://www.humanconnectome.org/documentation/S500} \cite{McNab2013}, containing the HARDI MRI data of healthy human subjects between 22 and 35 years of age. 

\subsection*{Construction of the Graphs}

The workflow for generating the connectomes has applied the CMTK toolkit \cite{Daducci2012}, including the FreeSurfer tool and the MRtrix tractography processing tool \cite{Tournier2012} with randomized seeding and with the deterministic streamline method, with 1 million streamlines. For the present study we have applied the 463-vertex graph resolution. 

The parcellation data (containing the ROI labels) is listed in the CMTK nypipe GitHub repository \url{https://github.com/LTS5/cmp_nipype/blob/master/cmtklib/data/parcellation/lausanne2008/ParcellationLausanne2008.xls}.

 The further details of braingraph-constructions are given in \cite{Kerepesi2016b}. 
 
 The braingraphs can be downloaded from the \url{https://braingraph.org/cms/download-pit-group-connectomes/} site, by choosing the ``Full set, 413 brains, 1 million streamlines'' option. In the present study we have used the 463-node resolution graphs.

\subsection*{The Apriori Algorithm}

The apriori algorithm \cite{Agrawal1994a} is a well-known tool in data mining for selecting the frequent item sets from a large collection of subsets of a big item set. In constructing association rules \cite{Assoc,Agrawal1994} the selection of frequent subsets is a basic step of the rule construction. In general, an $n$ element set has $2^n$ subsets, therefore for not-too-small $n$'s it is not feasible to review {\em all the} $2^n$ subsets. However, if we want to identify only the subsets with high enough frequency (or ``support'', in data mining terms \cite{han-kamber}), then we can make use of the following observation: Suppose that the set $A$ has frequency $\alpha\geq 0$. Then all subsets of $A$ has a frequency at least $\alpha$. Therefore, first, we identify those 1-element subsets, which has a frequency at least $\alpha$, it is an easy task. Then, for identifying all the 2-element subsets of frequency $\alpha$ we need to screen only the pairs of the frequent 1-element subsets. Next, for identifying the frequent 3-element subsets, we take all the frequent 2-element subset appears with exactly one common element, and verify if their union is frequent or not. The algorithm is continued in a similar way, and usually, it finishes quickly.

Here we have applied an adaptation of the apriori code from the website \url{http://adataanalyst.com/machine-learning/apriori-algorithm-python-3-0/} with small modifications.

\subsection*{Statistical Analysis}

For a chosen frequent neighbor set $F$, we have counted its occurrences in the male dataset by $count_1(F)$ and in the female dataset by $count_2(F)$.
The support was calculated as $supp_i(F) = \frac{count_i(F)}{S_i}$, where $S_i$, for $i=1,2$, wh is the number of male and female braingraphs, respectively.
For each set $F$ we need to determine whether $supp_1(F)$ and $supp_2(F)$ significantly differ.
For this goal we used the chi-squared test for categorical data:

\begin{center}
	\small
	\begin{tabular}{ l | c | c | c }
		& contains $F$ & does not contain $F$ & total \\ 
		\hline
		1$^{st}$ sample & $count_1(F)$ & $S_1 - count_1(F)$ & $S_1$ \\  
		2$^{nd}$ sample & $count_2(F)$ & $S_2 - count_2(F)$ & $S_2$ \\  
		\hline
		total & $count_1(F) + count_2(F)$ & $S_1+S_2-count_1(F)-count_2(F)$ & $S_1 + S_2$   
	\end{tabular}
\end{center}

Then the test is calculated as 
\[ \chi^2 = \frac{(count_1(F) \cdot (S_2-count_2(F))-count_2(F) \cdot (S_1-count_1(F)))^2 \cdot (S_1+S_2)}{S_1 \cdot S_2 \cdot (count_1(F)+count_2(F)) \cdot(S_1+S_2-count_1(F)-count_2(F))}\]

The degree of freedom for this test is one because it is the number of samples minus one times the number of categories minus one.

\noindent {\bf Holm-Bonferroni correction \cite{Holm1979}:} After computing the $p$ value for every frequent set,  we ordered these p-values $p_1 \leq p_2 \leq p_3 \leq \ldots \leq p_m$.
With a significance level $\alpha = 0.01$ let the Holm-Bonferroni value for $k^{th}$ frequent set be $p_k^{'} = \frac{\alpha}{m+1-k}$.
Then let $t$ be the minimal index such that $p_t > p_t^{'}$: we have to reject the null hypotheses for $i$ indices $i\leq t$.
If $t=1$, then we do not reject any null hypotheses thus the difference in supports is significant.

\section*{Conclusions}

First in the literature, we have mapped the frequent neighbor sets of the human hippocampus, by applying the Frequent Network Neighborhood Mapping method. We have identified the frequent neighbor sets of the human hippocampus, and we have also compared the data of healthy young men and women in respect to the neighborhood of the hippocampus. We have found that men have much more significantly more frequent neighbor sets of the hippocampus than women. Our results are in contrast with the generally much better connection properties of the braingraphs of women than of men, as reported in \cite{Szalkai2015,Szalkai2016a,Szalkai2015c}. Our results also need to be compared to the volumetric studies \cite{Tan2016}, where it was shown that hippocampal volumes, relative to the intracranial volume and also to the total brain volume are the same in the two sexes. Therefore, we have shown that the neighbors of the human hippocampus significantly differ in men and women, while there are no relative volumetric differences in the hippocampus.

\section*{Data availability} The data source of this study is Human Connectome Project's website at \url{http://www.humanconnectome.org/documentation/S500} \cite{McNab2013}. 

The parcellation data, containing the ROI labels, is listed in the CMTK nypipe GitHub repository \url{https://github.com/LTS5/cmp_nipype/blob/master/cmtklib/data/parcellation/lausanne2008/ParcellationLausanne2008.xls}.

The braingraphs can be downloaded from the \url{https://braingraph.org/cms/download-pit-group-connectomes/} site, by choosing the ``Full set, 413 brains, 1 million streamlines'' option. In the present study we have used exclusively the 463-node resolution graphs.

Supplementary Tables S1 and S2 contain the one, two three and four-element subsets of the neighbor-sets of the left- and the right hippocampi, respectively, with a minimum frequency of 0.9. The Supplementary Table S3 lists those neighbor-sets of the left hippocampus, which are significantly more frequent in female connectomes; while Table S4 lists those, which are significantly more frequent in male connectomes. Similarly, Tables S5 and S6 contain the analogous data for the right hippocampus. The supplementary tables can be downloaded in a compressed MS Excel format from \url{http://uratim.com/hpc/supplementary_HCP_neighbors.zip}.

\section*{Acknowledgments}
Data were provided in part by the Human Connectome Project, WU-Minn Consortium (Principal Investigators: David Van Essen and Kamil Ugurbil; 1U54MH091657) funded by the 16 NIH Institutes and Centers that support the NIH Blueprint for Neuroscience Research; and by the McDonnell Center for Systems Neuroscience at Washington University. VG and BV were partially supported by the VEKOP-2.3.2-16-2017-00014 program, supported by the European Union and the State of Hungary, co-financed by the European Regional Development Fund, VG and MF by the NKFI-126472 grant of the National Research, Development and Innovation Office of Hungary. BV and MF was supported in part by the EFOP-3.6.3-VEKOP-16-2017-00002 grant, supported by the European Union, co-financed by the European Social Fund.
\bigskip 

\noindent Conflict of Interest: The authors declare no conflicts of interest.

\section*{Author Contribution} MF analyzed braingraphs and developed statistical and graph theoretical tools, BV constructed the image processing system and computed the braingraphs, VG has initiated the study, analyzed data and wrote the paper.



\end{document}